# Developer Perceptions and Adoption in AI-Assisted Software Development

Mark Looi[1]

[1] Mark Looi, 3828 Cascadia Avenue South, Ste 101, Seattle, WA 98118. +1 425 941 2378. marklooi@looiconsulting.com.

Developer Perceptions and Adoption in AI-Assisted Software Development 1

# Abstract

The use of AI-assisted development tools has raised questions about their relationship to developer productivity, software quality, and adoption behavior. This study empirically explores AI tool usage, perceived productivity, perceived code quality, and future adoption intent using survey data from 147 professional software developers.

The findings indicate that broader and more frequent AI tool usage is associated with higher perceived productivity, while broader usage is also associated with higher perceived code quality and stronger future adoption intent. Perceived productivity and perceived code quality are themselves positively associated, suggesting that developers perceive no trade-off between speed and quality.

The results identify differences between coding- and testing-related AI usage. AI tool usage in testing is less extensive than in coding, with lower reported productivity gains.

Regression analysis indicates that future adoption intent is positively associated with frequency of AI tool usage and ease of workflow integration, while security concerns exhibit a modest negative association.

Clustering analysis identifies three developer segments, Enthusiasts, Pragmatists, and Cautious, which differ in AI tool usage, perceived productivity, perceived code quality, future outlook, and adoption intent. The identified segments are consistent with diffusion-oriented patterns of technology adoption where more highly engaged developers report broader AI tool usage and higher levels of organizational AI policy presence.

# 1 Introduction

The use of Artificial Intelligence (AI) tools in software development has expanded rapidly in recent years, prompting growing interest in how developers integrate these tools into software engineering practice.

This study employs an exploratory empirical approach to examine developers' reported use of AI tools in software development, focusing on perceived productivity, perceived code quality, and future adoption intent. The analysis compares usage patterns across coding and testing activities and examines relationships among AI tool usage, perceived outcomes, and future adoption intent. Because coding and testing activities involve different workflows, verification, and reliability requirements, developers may perceive and use AI tools differently across these contexts.

This study analyzes survey responses from 147 professional developers regarding AI tool usage, perceived productivity, perceived code quality, and future adoption intent. Given the perception-based and cross-sectional nature of the data, the study focuses on identifying patterns of association rather than establishing causal relationships.

The findings indicate that developers report broader and more frequent AI tool usage in coding-related activities than in testing-related activities, suggesting a testing gap in usage. More frequent and broader AI tool usage are positively associated with perceived productivity and future adoption intent, while perceived productivity and perceived code quality tend to co-occur in developers' experience. The study also identifies distinct developer engagement profiles characterized by differing levels of AI usage, perceived usefulness, and future adoption intent.

This paper makes the following contributions:

- An empirical analysis of developers' AI tool usage patterns across coding and testing activities
- Evidence that broader and more frequent AI tool usage is associated with higher perceived productivity, higher perceived code quality, and stronger future adoption intent
- Evidence of a testing gap in AI tool usage, with testing-related AI usage reported as less extensive and less productive than coding-related usage
- A characterization of developer engagement profiles (Enthusiasts, Pragmatists, and Cautious) associated with differing patterns of AI usage, perceived outcomes, and future adoption intent

# 2 Background and Literature Review

This literature review situates the study within prior research on AI-assisted software development, developer perceptions of AI tools, and technology adoption in software engineering practice.

## 2.1 Empirical Studies of AI Tool Use in Software Engineering

Research on AI-assisted software development has examined both software outcomes and developer experiences. Objective studies have reported mixed findings regarding the effects of AI-assisted development on productivity and code quality (He et al., 2025; Becker et al., 2025).

Survey-based studies examine how developers perceive and use AI tools in practice. Prior work reports that developers often perceive AI tools as beneficial for productivity and workflow support, while also expressing concerns related to correctness, trust, and integration into existing workflows (Liang et al., 2024; Russo, 2024; Sergeyuk et al., 2025; Vaithilingam et al., 2022).

Together, these studies illustrate that AI-assisted software development can be examined through both software artifacts and developer experiences. The present study focuses on the latter perspective by examining how developers report using AI tools in practice and how those experiences relate to perceived outcomes and future adoption intent.

## 2.2 Technology Adoption and Perceived Utility

Research on technology adoption emphasizes the importance of perceived usefulness and perceived ease of use in shaping system adoption behavior (Davis, 1989; Venkatesh et al., 2003). In AI-assisted software development contexts, developers' perceptions of usefulness, reliability, and workflow fit may influence both current usage patterns and future adoption intent.

Prior work has also identified concerns related to AI-assisted development, including security vulnerabilities and intellectual property risks (Negri-Ribalta et al., 2024). These concerns may shape how developers evaluate and integrate AI tools into software development workflows.

These perspectives motivate examining how developers' experiences with AI tools relate to usage patterns and future adoption intent in software development practice.

### 2.3 Adoption Frameworks and Developer Behavior

Diffusion of innovations theory has also been used to describe how technology adoption varies across populations and evolves over time (Rogers, 2003). In AI-assisted development contexts, prior work highlights the potential influence of organizational context, developer cognition, and workflow integration on adoption behavior (Russo, 2024).

These frameworks help interpret differences in developer engagement and adoption behavior.

### 2.4 Study Focus

This study examines how developers report using AI tools across software development activities, with particular attention to differences between coding- and testing-related usage. The analysis focuses on perceived productivity, perceived code quality, and future adoption intent, examining relationships among usage patterns, perceptions, and adoption-related measures using survey data from professional developers.

Given the perception-based and cross-sectional nature of the data, the study focuses on identifying patterns of association rather than establishing causal relationships.

## 3 Research Questions (RQs)

### 3.1 Current State of AI Use

RQ1: What are the patterns of developers' AI tool usage across coding and testing?

### 3.2 Usage and Perception

RQ2: How is AI tool usage associated with perceived productivity and perceived code quality?

### 3.3 Perception and Increased Usage

RQ3: How are perceived productivity and perceived code quality associated with developers' intent to increase AI tool usage?

### 3.4 Contextual Factors and Adoption Intent

RQ4: How are contextual factors such as integration effort, security concerns, and organizational policy associated with future AI tool adoption intent?

### 3.5 Segmentation

RQ5: Do distinct developer segments emerge based on AI usage, perceived outcomes, and adoption patterns?

# 4 Methodology

## 4.1 Research Context

This study examines developers' reported use of AI tools in software development practice, with a focus on perceived productivity, perceived code quality, and future adoption intent. The analysis uses developers' experiences and perceptions rather than objective measurements of software artifacts or development outcomes.

## 4.2 Survey Design and Measures

Data was collected using a structured survey of 34 questions, including several multi-item and matrix questions capturing multiple variables.

Most perceptual items were measured using 5-point Likert scales (full question wording and response options are provided in the supplementary material, "Survey of AI in Software - Google Forms.pdf"). The survey captured:

- AI tool usage (frequency and breadth across coding and testing tasks)
- Perceived productivity in coding
- Perceived productivity in testing
- Perceived code quality
- Adoption intent, measured as likelihood of increased AI usage
- Contextual factors, including integration effort, cost, security, and policy considerations

## 4.3 Variables and Indices

To reduce dimensionality and improve reliability, related survey items were aggregated into composite indices. Indices were calculated using either summation or averaging of Likert-scale responses, depending on the construct. The primary indices used in the analysis were:

- Intent to Increase Usage Index (dependent variable)
- Perceived Quality Index (PQI)
- AI Coding Tool Index (breadth of coding tool usage)
- AI Testing Tool Index (breadth of testing tool usage)
- Mean Concern Index (perceived challenges associated with AI-assisted development)

Internal consistency of composite measures was assessed using Cronbach's alpha, with values indicating acceptable reliability for exploratory research (Taber, 2018). For more details on the composition of the indices, see Appendix Table 2: Composite Indices Definitions.

## 4.4 Data Collection and Participants

The survey was distributed to software professionals through email and social media channels associated with the author during October and November 2025. A total of 147 valid responses from individuals actively engaged in software development activities were included in the analysis.

Participants represented a range of roles, experience levels, and geographic regions (see Supplement Appendix Table 1: Survey Demographics). Participation in the survey was voluntary and no personally identifiable information was collected. Respondents could choose not to participate or discontinue at any time. Completion of the survey was taken as implied consent, consistent with minimal-risk research practices.

The full survey instrument and cleaned dataset are provided in the supplementary material.

### 4.5 Statistical Analysis

Given the ordinal nature of the survey data, Spearman rank correlation **($\rho$)** was used to examine associations between variables. The research questions were addressed as follows:

- RQ1: descriptive statistics were used to characterize AI tool usage patterns across coding and testing activities.
- RQ2–RQ3: Spearman correlations were used to examine associations among AI tool usage, perceived outcomes, and future adoption intent.
- RQ4: linear regression was used to examine associations between contextual factors and future adoption intent.
- RQ5: principal component analysis (PCA) followed by K-means clustering was used to identify potential developer segments

The number of clusters ($k = 3$) was selected based on inspection of the within-cluster sum of squares curve and interpretability of the resulting clusters. Silhouette analysis indicated slightly higher scores for $k = 2$, but this solution produced overly coarse groupings. The $k = 3$ solution provided a better balance between cluster separation and interpretability. See 12.1.

Statistical significance was evaluated at $\alpha < 0.05$.

## 5 Findings

### 5.1 RQ1: Current AI Usage

AI tools are widely used across software development activities, particularly in coding.

Sixty-seven percent of respondents report using AI coding tools “often” or “always”, with high perceived utility for coding (Table 1). Supporting details are provided in the Supplement, Figure 1.

Usage breadth differs across domains. Developers engage in a median of 5 coding-related AI activities (out of 11), while testing usage is more limited (median = 2 activities).

Perceived utility of AI tools is generally high, with more favorable ratings for coding than testing activities (Table 1).

| Metric | Area | Mean (1-5) | % "High/Very High" |
|---|---|---|---|
| Accuracy/Relevance | Coding | 4.17 | 53% |
| Bug-Finding Effectiveness | Testing | 3.36 | 43% |

*Table 1: Developer Ratings of Utility*

Usage patterns for coding activities show both substantial usage and variation in breadth (Figure 1). The median developer engages in 5 activities, with nearly half (49%) using 6 or more, while 23% engage in 3 or fewer. Code completion is the most common use case (67%), see Supplement Figure 2.

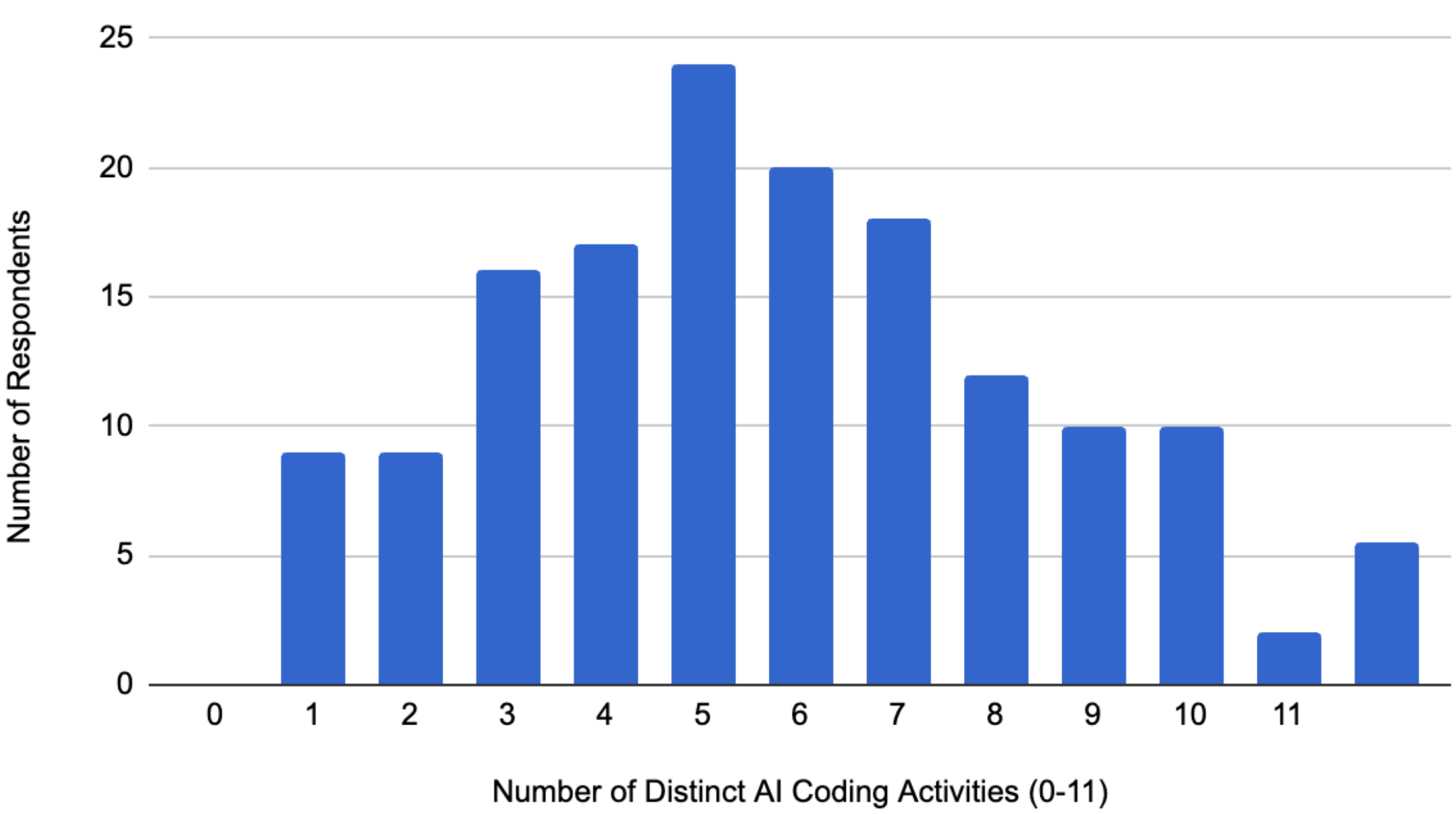

*Figure 1: Distribution of AI Coding Activities*

In contrast, testing usage is more limited and concentrated (Figure 2). While 68% of developers report using AI testing tools and 43% say they use them "Often" or "Always", the mean number of activities is 2.29 (median = 2). Over one-third (35%) engage in one or no testing-related AI activities, although 38.9% report using three or more. Test case generation is the most common testing use case (72%), followed by test automation (60%), while other activities are less widely used. Supporting details are provided in the Supplement, Figures 6 and 7.

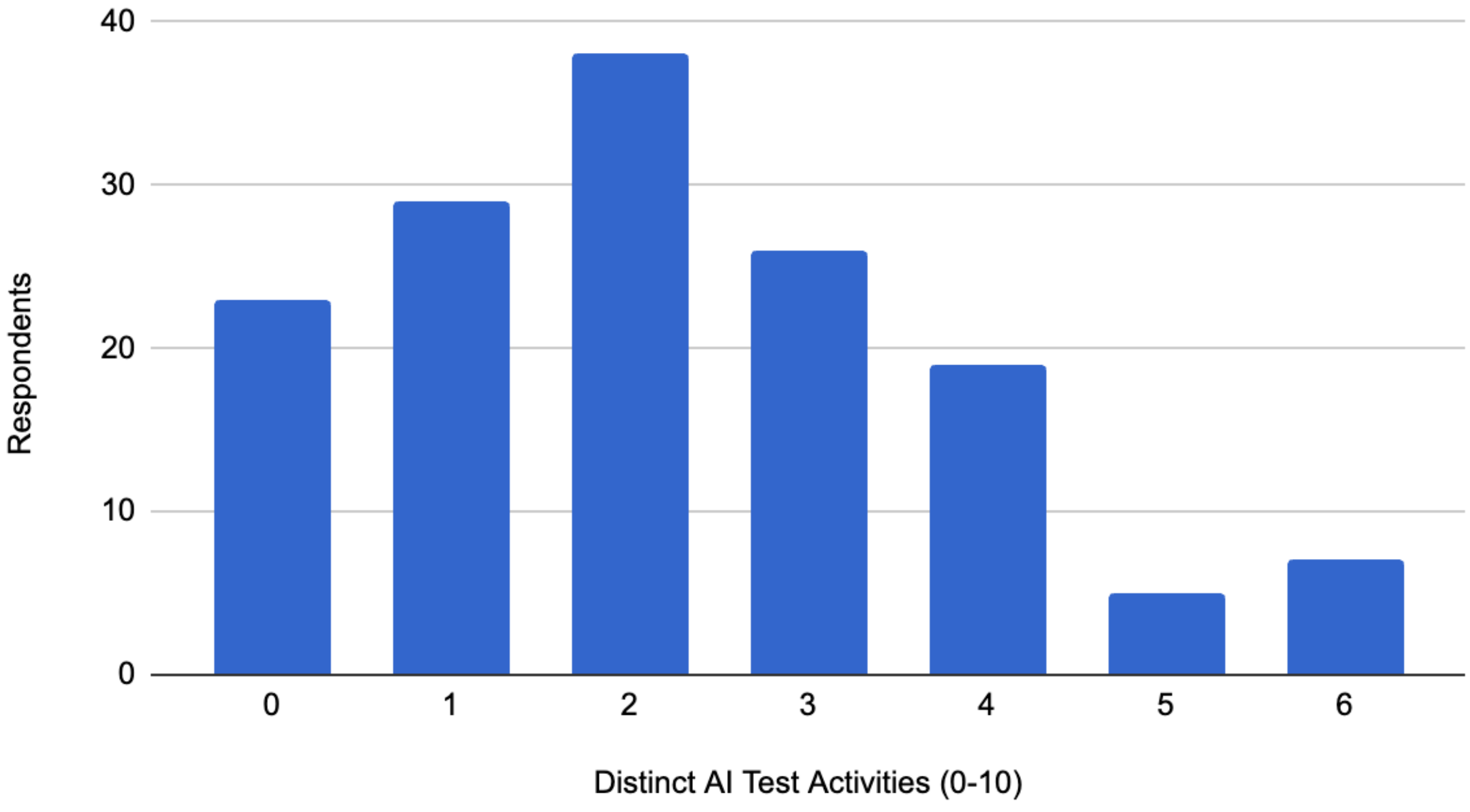

*Figure 2: Distribution of AI Testing Activities*

We next examine the association of usage to perceived productivity gains and quality.

## 5.2 RQ2: Associations with Perceived Productivity and Quality

AI tool usage is positively associated with perceived productivity in both coding and testing contexts, with stronger relationships observed for coding activities. Developers report higher productivity gains for coding than for testing (Table 2) with the median for coding twice that of test.

| Perceived Productivity Measure | % Reporting > 3 Hours Saved | Mean (1–6) | Median (1–6) |
|---|---|---|---|
| Coding activities | 60% | 3.92 | 4 |
| Testing activities | 48% | 3.14 | 2 |

*Table 2: Perceived Productivity (PP)*

Most developers report that AI tools have a positive impact on perceived code quality. Approximately 74% of respondents report improvements in code quality, while 68% report improvements in testing contexts. Details are provided in the Supplement, Figures 5 and 10.

Correlation analysis shows that both frequency and breadth of AI tool use are positively associated with perceived productivity in coding and testing (Table 3). Frequency of use exhibits the strongest associations, while breadth of use (as captured by the AI Coding Tool Index and AI Testing Tool Index) is also positively related to perceived productivity.

| Variable Pair | N | ρ | p-value |
|---|---|---|---|

| AI Coding Tool Index – Perceived productivity code | 145 | 0.347 | < 0.001 |
|---|---|---|---|
| Frequency of AI coding tool use – Perceived productivity code | 147 | 0.458 | < 0.001 |
| AI Testing Tool Index – Perceived productivity test | 132 | 0.357 | < 0.001 |
| Frequency of AI testing tool use – Perceived productivity test | 132 | 0.479 | < 0.001 |
| PQI – Perceived productivity code | 147 | 0.190 | 0.021 |
| PQI – Perceived productivity test | 132 | 0.358 | < 0.001 |

*Table 3: Perceived Productivity Findings (Note: N varies because not all Developers used AI testing tools.)*

Perceived productivity is also positively associated with perceived code quality, with a weaker association observed for coding-related activities and a stronger association observed for testing-related activities.

Higher breadth of AI tool usage is also positively associated with perceived code quality. As shown in Table 4, both coding and testing tool breadth exhibit weak but statistically significant correlations with PQI.

| Variable Pair | N | ρ | p |
|---|---|---|---|
| AI Coding Tool Index – PQI | 132 | 0.2356 | 0.00407 |
| AI Testing Tool Index – PQI. | 132 | 0.289 | 0.00039 |

*Table 4: Breadth and Perceived Quality Findings*

Overall, the findings for RQ2 indicate that broader and more frequent AI tool usage is positively associated with perceived productivity, while perceived productivity and perceived code quality are also positively associated. These relationships are stronger in coding-related activities than in testing-related activities.

We next examine perceived quality and productivity and their association with adoption.

## 5.3 RQ3: Associations Between Perceived Outcomes and Adoption Intent

Developers report a high overall intent to increase their use of AI tools, particularly for coding (Table 5).

| Category | N | Mean | Median |
|---|---|---|---|
| Coding Tools | 147 | 4.32 | 5 |
| Testing Tools | 145 | 3.89 | 4 |
| Composite Index | 147 | 4.08 | 4 |

*Table 5: Intent to Increase Usage of AI Tools*

Perceived productivity, perceived code quality, and breadth of AI tool usage are all positively associated with future adoption intent (Table 6).

| Variable Pair | N | ρ | p-value |
|---|---|---|---|
| Perceived productivity in testing – Intent | 132 | 0.319 | < 0.001 |

| Perceived productivity in coding – Intent | 147 | 0.227 | 0.006 |
|---|---|---|---|
| PQI – Intent | 147 | 0.231 | 0.005 |
| AI Coding Tool Index – Intent | 147 | 0.317 | < 0.001 |
| AI Testing Tool Index – Intent | 132 | 0.347 | < 0.001 |

*Table 6: Associations with Adoption Intent*

Overall, the findings for RQ3 indicate that future adoption intent is positively associated with perceived productivity, perceived code quality, and broader AI tool usage. The strongest associations are observed for testing-related productivity and testing-related breadth of usage.

## 5.4 RQ4: Associations with Contextual Factors and Adoption Intent

To examine associations between contextual factors and future adoption intent, a linear regression model was estimated using the Intent to Increase Usage Index as the dependent variable. The model included:

- frequency of AI testing tool use
- perceived quality (PQI)
- Ease of integration
- Cost
- Security concerns
- Organizational policy
- Developer interest

The model explains 29.5% of the variance in adoption intent ($R^2 = 0.295$, $F(7, 138) = 8.14$, $p < 0.001$).

| Variable | β | p-value |
|---|---|---|
| Frequency of AI testing tool use | 0.38 | < 0.001 |
| Ease of integration | 0.29 | 0.001 |
| Security concerns | -0.17 | 0.001 |
| Cost | n.s. | |
| PQI | n.s. | |
| Organizational policy | n.s. | |
| Developer interest | n.s. | |

*Table 7: Regression Results Predicting Adoption Intent*

Frequency of AI tool use and ease of integration are positively associated with adoption intent, while security concerns are negatively associated. Cost, perceived quality, organizational policy, and developer interest are not statistically significant in this model.

In this model, tool usage and integration are more strongly associated with adoption intent than broader contextual or organizational factors.

## 5.5 RQ5: Developer Segmentation

To examine how usage, perceived outcomes, and adoption-related measures relate across developers, a clustering analysis was performed using four variables: Intent to Increase Usage, AI Coding Tool Index (usage breadth), Strategic Outlook Index, and the Perceived Quality Index (PQI). Principal component analysis (PCA) was applied for dimensionality reduction, followed by K-means clustering. A three-cluster solution was selected based on interpretability and stability, although alternative segmentations are plausible.

The analysis identifies three developer segments:

- Enthusiasts (n = 56): High usage breadth, higher perceived productivity and perceived quality, and high intent to increase usage.
- Pragmatists (n = 53): Moderate usage and perceived outcomes, with generally positive but less pronounced intent and outlook.
- Cautious (n = 38): Lower usage breadth, weaker perceived outcomes, and lower intent to increase usage.

Enthusiasts tend to report higher agreement regarding the importance of emerging AI-related skills and future AI development patterns, while Cautious developers report lower levels of agreement across these measures. Pragmatists occupy an intermediate position in usage breadth and strategic outlook, although their perceived quality ratings remain notably lower than those of Enthusiasts and similar to the Cautious.

The clustering variables capture approximately 42% of the variance in the data, suggesting that these dimensions account for a substantial portion of the variation in developer perceptions and usage patterns.

| | Enthusiasts | Cautious | Pragmatists | Total/Average |
|---|---|---|---|---|
| Number | 56 | 38 | 53 | 147 |
| Intent to Increase Usage Index | 4.49 | 2.87 | 4.51 | |
| Strategic Outlook Index | 3.77 | 2.74 | 3.77 | |
| AI Coding Tool Index (Usage Breadth) | 7.12 | 3.74 | 5.21 | |
| PQI | 4.69 | 3.66 | 3.40 | |
| Mean Concern Index Score (1-5) | 3.18 | 3.46 | 3.22 | 3.27 |
| AI Coding Policy (Yes) | 33 | 2 | 14 | 49 |
| AI Coding Policy (No) | 23 | 36 | 39 | 98 |
| Policy Support % | 58.9% | 5.3% | 26.4% | 33.3% |

*Table 8: Developer Archetypes: Enthusiasts, Cautious, Pragmatists Clusters*

To further explore organizational context, the presence of AI-related development policies was compared across clusters. A chi-square test indicates a significant association between cluster membership and policy presence ($\chi^2 = 45.59$, df = 2, $p < 0.001$). Enthusiasts are more likely to report the presence of organizational AI coding policies (59%) compared to Pragmatists (26%) and Cautious developers (5.3%).

This association suggests that organizational context and developer segmentation are related, although the direction of this relationship cannot be determined from the present data.

Overall, the segmentation suggests that Enthusiasts stand apart through broader AI tool usage, higher perceived productivity and quality, and stronger future adoption intent. Pragmatists appear engaged but more critical in their evaluation of AI-assisted development, while the Cautious group exhibits lower engagement and weaker expectations regarding AI tool usage.

## 5.6 Developer Perspectives on Future AI Development

Developers report a generally positive but nuanced outlook on AI-assisted software development, characterized by strong expectations for continued integration alongside recognition of technical and organizational challenges.

In terms of future capabilities, respondents assign relatively high likelihood to applications that anticipate user needs (mean = 4.06) and incorporate multimodal interaction (mean = 3.83). More dynamic interface paradigms, such as fully conversational or AI-generated user interfaces, receive comparatively lower, but still moderate, levels of agreement (see Appendix Table 3).

Respondents also report that emerging architectural patterns, such as retrieval-augmented generation and agentic systems, are of moderate to high importance (see Appendix Table 4). Developers also identify several challenges associated with AI-native systems. Model hallucinations, security risks, and testing complexity are all rated as challenges (see Appendix Table 5).

Skills related to orchestration, data management, and prompt engineering are consistently rated as important (Table 9).

| Future Developer Skill Importance | N | Mean | Median |
|---|---|---|---|
| Importance of Orchestration Workflow Design | 142 | 3.98 | 4 |
| Importance of Data management | 142 | 3.9 | 4 |
| Importance of Prompt Engineering | 144 | 3.9 | 4 |
| Importance of Model Selection | 142 | 3.54 | 4 |
| Importance of Foundation model fine tuning | 143 | 3.41 | 3 |

*Table 9: Future Skills*

Strategic outlook is positively associated with several factors, including the perceived importance of orchestration ($\rho = 0.406$), AI-native development status ($\rho = 0.427$), and intent to increase usage ($\rho = 0.305$), as shown in Table 10. Associations are also observed with perceived productivity measures, although these are weaker in magnitude.

| Variable Pair | N | ρ | p-value |
|---|---|---|---|
| Importance of Orchestration – Strategic Outlook | 142 | 0.406 | < 0.001 |
| Importance of Prompt Engineering – Strategic Outlook | 144 | 0.320 | < 0.001 |
| AI-Native Development Status – Strategic Outlook | 144 | 0.427 | < 0.001 |

| Perceived productivity in coding – Strategic Outlook | 145 | 0.242 | 0.003 |
|---|---|---|---|
| Perceived productivity in testing – Strategic Outlook | 132 | 0.249 | 0.004 |
| Intent to Increase Usage – Strategic Outlook | 147 | 0.305 | < 0.001 |

*Table 10: Associations with Strategic Outlook*

# 6 Discussion

## 6.1 Usage and Perceived Productivity (RQ1–RQ2)

The findings indicate that AI tools are widely used across software development activities, particularly in coding, with lower but still substantial use in testing. Usage spans multiple tasks, although the breadth of application differs across coding and testing contexts.

AI tool usage is positively associated with perceived productivity in both coding- and testing-related activities. These associations are stronger for coding activities, where respondents report greater time savings and broader tool usage. A notable gap exists between coding and testing in both usage breadth and perceived productivity. While many developers report engaging with multiple coding-related AI activities, testing usage tends to be concentrated in fewer tasks.

These findings are broadly consistent with prior work reporting perceived productivity gains from AI-assisted development (Bird et al., 2023; Liang et al., 2024; Vaithilingam, 2022). However, the present study focuses on developers' perceptions and experiences rather than objective productivity outcomes. The results therefore should not be interpreted as evidence of objective productivity improvements or declines.

## 6.2 Perceived Productivity and Quality (RQ2)

The findings show that usage breadth and frequency of AI tool use are positively associated with perceptions of increased productivity in both coding and testing. Perceived productivity and perceived code quality are also positively associated. Together, these relationships suggest that broader usage, higher perceived productivity, and higher perceived quality tend to co-occur in developers' experience.

This pattern differs from objective studies reporting potential quality concerns (He et al., 2025) or productivity trade-offs in AI-assisted development (Becker et al., 2025). The present findings do not directly support or contradict those studies, as this study examines developers' perceptions rather than objective software outcomes.

The distinction between objective outcomes and developer perceptions remains important because perceived usefulness is a primary determinant of technology adoption and system use (Venkatesh et al., 2003). In the current study, developers who report broader AI tool usage also tend to report more favorable perceived outcomes overall, including higher perceived productivity and perceived code quality. These findings align with Venkatesh et al. (2003) and suggest that these perceptions may support continued engagement with AI tools in practice, regardless of whether objective quality outcomes align with these perceptions.

## 6.3 Adoption Intent and Contextual Factors (RQ3–RQ4)

Developers report a high intention to increase their future use of AI tools, particularly for coding-related activities. Future adoption intent is positively associated with broader AI tool usage, perceived productivity, and perceived code quality. Together, these findings suggest that developers who use AI tools more extensively and perceive greater productivity and quality benefits are also more likely to intend continued AI tool adoption.

Regression analysis further indicates that certain contextual factors are associated with adoption intent when considered jointly. Frequency of use and ease of integration show positive associations with intent, while security concerns show a small negative association. Other factors, including cost and organizational policy, do not show statistically significant associations in this model.

These findings are broadly consistent with technology adoption research emphasizing perceived usefulness and ease of use as important correlates of adoption behavior (Davis, 1989; Venkatesh et al., 2003). The results suggest that developers' experiences using AI tools may be more closely associated with future adoption intent than broader organizational or contextual considerations in this sample.

## 6.4 Developer Segmentation (RQ5)

The clustering analysis identifies heterogeneity in developers' AI tool usage, perceived outcomes, and future adoption intent. The identified segments—Enthusiasts, Pragmatists, and Cautious developers—differ in usage breadth, perceived productivity, perceived quality, and intended future AI tool usage (Table 8).

These patterns are broadly consistent with diffusion-oriented perspectives on technology adoption, where different groups exhibit differing levels of engagement and adoption readiness (Rogers, 2003). Enthusiasts correspond roughly to Innovators and Early Adopters. Pragmatists approximate the Early Majority, adopting once evidence of successful use and organizational support become visible. The Cautious group shares characteristics with the Late Majority, exhibiting lower usage and more reserved expectations. However, the present segments are derived empirically from usage and perception variables.

The findings suggest that broader AI tool usage, higher perceived productivity, higher perceived quality, and stronger future adoption intent tend to co-occur across distinct but overlapping profiles of engagement with AI tools. These patterns are consistent with diffusion-oriented interpretations in which early, highly engaged users help demonstrate the utility of AI tools within teams, while organizational policies emerge alongside or in response to increasing adoption. Less engaged developers may be more likely to adopt once both peer usage and organizational support are visible.

The presence of organizational AI policies is also associated with segment membership, with higher prevalence among more engaged groups. However, the direction of this relationship cannot be determined from the present data.

While the present data does not allow causal or temporal ordering to be established, these patterns suggest that AI adoption within organizations may involve interactions between individual experience, peer effects, and organizational support.

## 6.5 Future Outlook

Developers generally report positive expectations regarding the future role of AI-assisted software development. Respondents assign relatively high importance to orchestration-related skills, data management, and prompt engineering, while also reporting interest in emerging architectural patterns such as retrieval-augmented generation and agentic systems (Table 9).

Strategic outlook is positively associated with factors such as engagement in AI-assisted development, perceived importance of orchestration-related skills, and future adoption intent. These relationships suggest that developers reporting broader AI tool engagement also tend to report more positive expectations regarding future AI development practices.

However, concern levels are relatively similar across the identified developer segments, suggesting that differences in future outlook are not strongly differentiated by perceived risks at the segment level.

The Enthusiast, Pragmatist, and Cautious segments also differ in their expectations regarding future AI capabilities, architectural patterns, and required skills (Table 8).

Overall, the future-outlook results extend the adoption findings by showing that developers' current AI engagement is linked not only to intended future use, but also to expectations about emerging skills, architectures, and development practices.

# 7 Implications of the Research

This study provides practical implications for both organizations and individual developers.

## 7.1 Implications for Development Teams and Management

### 7.1.1 Integration and Usage Patterns

The relationships among AI tool usage breadth, frequency of use, perceived productivity, perceived quality, and adoption intent suggest that organizations may benefit from supporting broader experimentation with AI tools across development activities rather than limiting usage to isolated tasks.

Ease of integration is also positively associated with future adoption intent. AI tools that align with existing workflows and reduce friction may therefore be more likely to support sustained developer engagement.

### 7.1.2 Skills and Architectural Orientation

Developers report relatively high perceived importance for orchestration-related skills, data management, prompt engineering, retrieval-augmented generation, and agentic systems. Developers may increasingly view AI-assisted software development as involving coordination of workflows, tools, and data sources in addition to direct interaction with language models.

Organizations may therefore consider aligning training and professional development efforts with these emerging areas.

### 7.1.3 Developer Segmentation

The identified developer segments suggest that AI engagement is not uniform across development teams. Organizations may therefore benefit from recognizing differences in developer experience, experimentation, and attitudes toward AI-assisted development rather than assuming a single adoption pattern across all users. More highly engaged developers can help demonstrate effective AI-assisted workflows and support knowledge sharing within teams, while less engaged developers may require greater organizational support, training, or opportunities for experimentation before broader adoption occurs.

These findings also suggest that organizational policies and governance structures may need to evolve alongside differing levels of AI engagement within teams.

#### 7.1.4 Adoption Context

The findings suggest that adoption behavior is more closely associated with workflow integration and developers' experiences using AI tools than with broader organizational or cost-related considerations. Organizations may therefore benefit from focusing on reducing friction in AI-assisted workflows and supporting integration into existing development practices.

Security concerns exhibit a modest negative association with future adoption intent, suggesting that clear governance, security guidance, and policy visibility may help support adoption among more cautious developers.

### 7.2 Implications for Developers and Professionals

#### 7.2.1 Usage and Workflow Integration

Higher frequency and broader use of AI tools are associated with more favorable perceptions of productivity and quality. Developers may therefore benefit from exploring AI-assisted workflows across a wider range of development activities, including documentation, testing, and code review in addition to code generation.

#### 7.2.2 Skill Development

The results indicate that developers place relatively high importance on prompt design, orchestration, and data integration. Developers may benefit from developing skills in these areas.

#### 7.2.3 Reflecting on Segmentation

The identified developer segments highlight that usage patterns and perceptions of AI tools vary. Understanding these archetypes may help developers reflect on their own level of engagement with AI-assisted development and guide decisions for skill development and tool adoption.

### 7.3 Further Research

Several avenues for future research emerge from the limitations and open questions identified in this analysis.

#### 7.3.1 Measurement of Productivity and Quality

The findings indicate a positive association between perceived productivity and perceived code quality. However, these measures are based on self-reported perceptions.

Prior research has examined objective measures of code quality and productivity in AI-assisted development, with some studies identifying potential trade-offs or declines in measured quality (e.g., He et al., 2025). The present study complements this work by focusing on developer perceptions, which do not reflect such trade-offs in developer experience.

Future research could jointly examine perceived and measured outcomes, exploring whether productivity improvements are accompanied by changes in defect rates, maintainability, or technical debt, and how these perceptions evolve over time. Longitudinal studies could help determine whether such discrepancies persist, diminish, or evolve as tools and development practices mature.

### 7.3.2 Temporal Dynamics and Work Reallocation

The cross-sectional design of this study does not capture how productivity gains evolve over time or how time savings are reallocated.

Future studies could use longitudinal methods or experience sampling approaches to examine productivity effects more precisely and to examine how developers use time saved through AI assistance. This could include investigating whether gains are offset by additional review, correction, or integration effort, or whether time is redirected toward higher-value activities such as system design or architectural work.

AI-assisted development may also complicate traditional approaches to measuring software engineering productivity and quality. Iterative prompting, conversational debugging, architectural exploration, and discarded generations may not be fully visible through repository-level artifacts or commit-based metrics alone. Future research combining perception-based methods, workflow telemetry, repository analysis, and longitudinal observation may provide a more comprehensive understanding of AI-assisted development practices.

### 7.3.3 Post-Adoption Effects and Hidden Costs

While perceived concerns such as cost and security show limited association with adoption intent in this study, post-adoption effects remain an open question.

Future research could examine whether AI-assisted development introduces additional forms of rework, such as verifying code provenance or checking correctness of generated code, and whether these offset perceived productivity gains. Understanding these dynamics would provide a more complete picture of the net impact of AI tools in practice.

### 7.3.4 A Gap between Coding and Testing

The observed difference in usage breadth between coding and testing suggests an area for further investigation. Future studies could examine the underlying causes of this imbalance, including differences in tool maturity, task complexity, organizational constraints, or developer confidence in testing-related AI tools. Understanding these factors may help explain variation in adoption across different stages of the development lifecycle.

### 7.3.5 Developer Segmentation and Adoption Trajectories

The identification of developer segments based on usage, perceptions, and outlook suggests a potential framework for understanding variation in adoption behavior.

Longitudinal research could examine how these segments evolve over time, including whether developers transition between segments as their experience with AI tools changes. Such studies could also assess whether segment membership is associated with differences in long-term adoption patterns or outcomes.

#### 7.3.6 Organizational Context and Policy

The association between organizational policy presence and developer segments raises questions about the role of governance in AI adoption.

Future research employing longitudinal or quasi-experimental designs could examine the directionality of this relationship. For example, studies could investigate whether policy development follows increased adoption, or whether formal policies influence subsequent usage patterns, or whether both are shaped by underlying organizational factors.

### 7.4 Threats to Validity and Limitations

This study provides empirical evidence on AI tool usage, perceived productivity, perceived quality, and developer outlook. However, several limitations should be considered when interpreting the findings relating to internal, external, and construct validity. As with all cross-sectional survey studies, the observed relationships should be interpreted as associative rather than causal.

#### 7.4.1 Internal Validity

Internal validity concerns the extent to which observed relationships reflect associations rather than confounding factors.

The study relies on self-reported measures (e.g., time saved, perceived quality, frequency of use). Such measures may be affected by recall bias, social desirability bias, or differences in individual interpretation. As a result, perceived productivity gains and quality assessments may not correspond to objective outcomes.

Independent and dependent variables were collected using the same survey instrument and from the same respondents. This may inflate observed associations due to shared method variance (Common Method Bias).

The observed associations do not establish directionality. For example, while higher AI usage is associated with higher perceived productivity, it is also possible that more experienced or more productive developers are more likely to adopt and use AI tools extensively.

#### 7.4.2 External Validity

External validity relates to the generalizability of the findings.

The study focuses on perceived productivity and quality (i.e., perception-based measures), which may differ from objective measures. As a result, findings may not directly translate to environments where objective performance metrics are the primary concern.

The sample (N = 147) was drawn from professional networks and voluntary participation, including industry contacts associated with the researcher, which may influence sample composition. While sufficient for statistical analysis, it may not fully represent the broader population of software developers. In particular, respondents may be more likely to have prior exposure to or interest in AI tools.

The AI development landscape is evolving rapidly. Perceptions of tools, risks, and architectural approaches may change over time, limiting the durability of the findings.

### 7.4.3 Construct Validity

Construct validity refers to how well the measures capture the intended concepts.

Constructs such as perceived productivity and perceived code quality are subjective developer assessments and were measured using ordinal scales. Differences in interpretation across respondents may introduce measurement error and affect the strength of observed associations.

Terms such as “agentic architecture,” “orchestration,” and “RAG” may not be uniformly understood by all respondents. This may introduce noise into measures related to strategic outlook and future expectations.

Some constructs (e.g., perceived impact of hallucinations) were measured using single survey items. Single-item measures may have lower reliability compared to multi-item scales and may not fully capture the complexity of the underlying concepts.

# 8 Conclusion

This study empirically examines AI tool usage, perceived outcomes, and future adoption intent in professional software development. The findings indicate that AI tools are widely used across software development activities, particularly in coding-related tasks, and that differences in usage patterns are associated with differences in perceived productivity, perceived code quality, and adoption intent.

Across the sample, broader and more frequent AI tool usage is associated with higher perceived productivity, while broader usage is also associated with higher perceived code quality and stronger future adoption intent. Perceived productivity and perceived code quality are positively associated, suggesting that developers do not experience a trade-off between speed and quality in their use of AI tools. These findings reflect developer perceptions and should be interpreted alongside prior work examining objective measures of software quality and productivity.

The results also highlight a difference between coding and testing contexts. While AI tool usage is widespread in coding, adoption in testing is less extensive, and perceived productivity gains are lower. AI-assisted workflows may evolve differently across stages of the development lifecycle.

Developers report positive expectations regarding the future role of AI-assisted software development, particularly in relation to orchestration, data integration, and emerging multi-model workflows, while also noting concerns about hallucinations, security, and testing complexity.

The analysis further identifies heterogeneity within the developer population. The Enthusiast, Pragmatist, and Cautious segments differ in AI tool usage, perceived productivity, perceived code quality, future outlook, and adoption intent. The findings are consistent with a diffusion-like interpretation in which more highly engaged developers report broader AI tool usage and higher levels of organizational AI policy presence.

Overall, the findings suggest that AI-assisted software development is shaped by the interaction of usage patterns, perceived outcomes, workflow integration, and organizational context. These factors characterize how developers experience and engage with AI-assisted software development practices.

# 9 Glossary

| Term / Acronym | Definition |
| --- | --- |
| **AI Coding Tool Index** | A composite index reflecting the breadth of AI coding tool usage across 11 specific coding activities (e.g., debugging, refactoring). |
| **AI Testing Tool Index** | A composite index reflecting the breadth of AI testing tool usage across 6 specific test activities (e.g., test cases). |
| **Testing Gap** | The observed disparity in AI tool usage and perceived productivity between testing and coding activities. |
| **Developer Archetypes** | Empirically derived clusters of developers based on usage patterns, perceived outcomes, and adoption intent (Enthusiasts, Pragmatists, Cautious). |
| **Perceived Quality Index (PQI)** | Composite measure of perceived code quality. |

*Table 11: Glossary*

# 10 Declarations

Funding: The authors did not receive support from any organization for the submitted work. No funding was received to assist with the preparation of this manuscript.

Data Statement: An anonymized dataset and data dictionary supporting the findings of this study are available at https://zenodo.org/records/20045102. Supplementary materials, including the survey instrument and additional figures, are provided with the submission. This study involved anonymous survey data and did not collect personally identifiable information. Participation was voluntary, and responses were used solely for research purposes. In accordance with institutional guidelines for minimal-risk research, formal ethical approval was not required.

Conflict of Interest: Mark Looi is an independent researcher and the President of Looi Consulting. The author has no relevant financial or non-financial interests to disclose.

During the preparation of this work the author used ChatGPT and Gemini in order to copy edit and construct tables. After using this tool/service, the author reviewed and edited the content as needed and takes full responsibility for the content of the published article.

# 12 Appendix

Additional supporting information including the survey instrument and responses are in the supplementary material, "Developer in Age of AI SUPPLEMENT".

## 12.1 Statistical Analysis Support for RQ5

### 12.1.1 Elbow Graph to Determine Number of Clusters

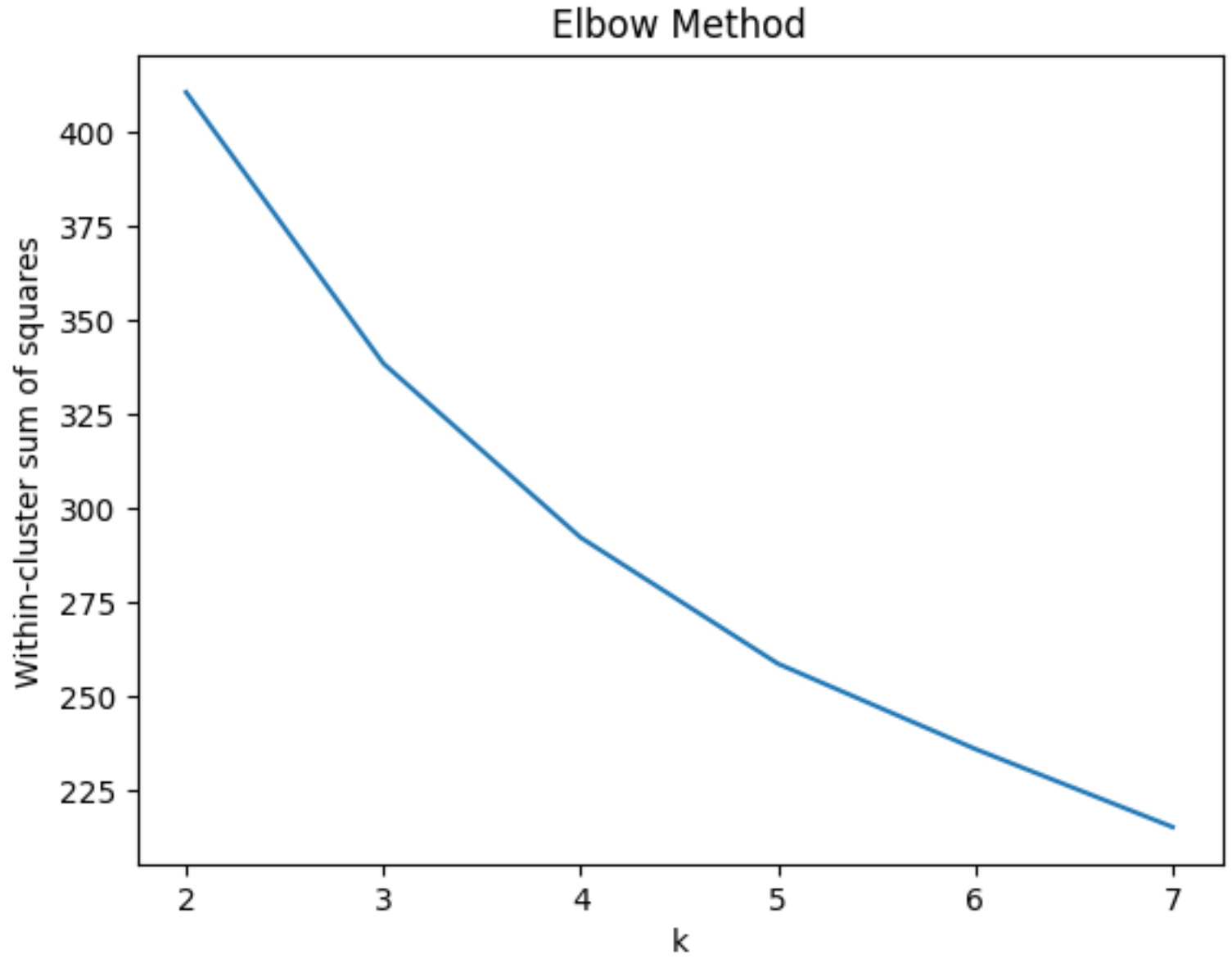

*Appendix Figure 1: Elbow Graph for RQ5*

Variables used in WSS to define profile space:

- Intent to Increase Usage Index
- Strategic Outlook Index
- AI Coding Tool Index
- Perceived Quality Index (PQI)

### 12.1.2 Cluster Validation Analysis

To further evaluate the clustering solution, silhouette analysis was performed for candidate K-means solutions with $k = 2$ through $k = 5$.

The silhouette analysis indicated slightly higher scores for the two-cluster solution. However, the $k = 2$ solution produced overly broad groupings that reduced interpretability and obscured meaningful differences in AI usage, perceived outcomes, and adoption intent. The three-cluster solution provided a better balance between cluster separation and interpretability, while solutions with $k \geq 4$ did not yield substantively meaningful additional structure.

These results support the use of the three-cluster solution reported in the study.

| Number of Clusters (k) | Silhouette Score |
|---|---|
| 2 | 0.281 |
| 3 | 0.232 |
| 4 | 0.228 |

| 5 | 0.234 |
|---|---|

*Appendix Table 1: Silhouette Scores for Candidate Cluster Solutions*

## 12.2 Additional Tables

These tables support sections in the paper.

| Index Name | Description | Components (Summary) | Construction |
|---|---|---|---|
| Intent to Increase Usage Index | Measures likelihood of increased AI use | Coding + testing intent items | Mean |
| Perceived Quality Index (PQI) | Perceived impact of AI on code quality | Coding perceived quality + testing perceived quality | Mean |
| AI Coding Tool Index | Breadth of AI use in coding | Count of coding-related AI activities | Sum |
| AI Testing Tool Index | Breadth of AI use in testing | Count of testing-related AI activities | Sum |
| Strategic Outlook Index | Forward-looking beliefs about AI-native development | Future UI, architecture, and skills items | Mean |
| Mean Concern Index | Perceived challenges of AI-assisted development | Security, hallucinations, testing, cost, etc. | Mean |

*Appendix Table 2: Composite Indices Definitions*

| Future UI Outlook | N | Mean | Median |
|---|---|---|---|
| Apps anticipate user needs | 145 | 4.06 | 4 |
| Apps use multimodal input | 145 | 3.83 | 4 |
| UI will be generated dynamically by AI | 144 | 3.6 | 4 |
| Primary UI will be conversational | 145 | 3.48 | 4 |

*Appendix Table 3: Future UI Outlook*

| Emerging Technology Importance | N | Mean | Median |
|---|---|---|---|
| RAG Relying on LLMs to call external APIs | 139 | 3.68 | 4 |
| Agentic Architecture | 138 | 3.62 | 4 |
| Hybrid Models (LLMs with models of physical world) | 137 | 3.55 | 4 |
| SLMs | 139 | 3.29 | 3 |
| Disposable Applications | 138 | 2.78 | 3 |

*Appendix Table 4: Emerging Architectural Patterns*

| Impact on AI-Native App Development | N | Mean | Median |
|---|---|---|---|

| AI models hallucinations | 140 | 3.84 | 4 |
|---|---|---|---|
| Security risks | 140 | 3.66 | 4 |
| Difficulty in testing | 141 | 3.6 | 4 |
| Cost | 141 | 3.51 | 4 |
| Lack of skilled developers | 141 | 3.39 | 3 |
| Risk of betting on obsolete technologies | 139 | 3.36 | 3 |
| Complexity | 140 | 3.29 | 3 |

*Appendix Table 5: AI App Development*